\documentclass[prl,aps,twocolumn,epsfig,superscriptaddress,nofootinbib]{revtex4}

\usepackage{amsmath}
\usepackage{amssymb}
\usepackage[dvips]{graphicx}
\usepackage{dcolumn}
\usepackage{bm}
\usepackage[colorlinks=false,dvipdfm]{hyperref} 
\usepackage{setspace}
\usepackage{amsfonts}
\usepackage{rotating}
\usepackage{makeidx}
\usepackage{amsthm}

\newcommand{\tr}{\mbox{Tr}}

\begin{document}

\title{Quantum cloning of steering \footnote{ Editors' Suggestion,  Chin. Phys. Lett. 39, 070302 (2022).\\ Reported in Physics World (https://physicsworld.com/a/cloning-quantum-steering-is-a-no-go/)}}

\author{Dian Zhu}
\affiliation{Department of Physics, School of Science, Tianjin University, Tianjin 300072, China}

 \author{Wei-Min Shang }
 \affiliation{Theoretical Physics Division, Chern Institute of Mathematics, Nankai University, Tianjin 300071, China}

\author{Fu-Lin Zhang\footnote{Corresponding author: flzhang@tju.edu.cn}}
\affiliation{Department of Physics, School of Science, Tianjin University, Tianjin 300072, China}

\author{Jing-Ling Chen \footnote{Corresponding author: chenjl@nankai.edu.cn}}
\affiliation{Theoretical Physics Division, Chern Institute of Mathematics, Nankai University, Tianjin 300071, China}

\date{\today}

\begin{abstract}
%
%
Quantum steering
 in a global state allows an observer to remotely steer a subsystem into different
ensembles by performing different local measurements on the other part.
%
%
We show that,
in general,
this property cannot be perfectly cloned by any joint operation between a steered subsystem and a third system.
Perfect cloning is viable if and only if the initial state is of zero discord.
We also investigate the process of cloning the steered qubit of a Bell state using a universal cloning machine.
Einstein-Podolsky-Rosen (EPR) steering, which is
a type of quantum correlation existing in the states without a local-hidden-state model,
is observed in the two copy subsystems.
This contradicts the conclusion of \emph{no-cloning of quantum steering (EPR steering)} [C. Y. Chiu \emph{et al.}, npj Quantum Inf. 2, 16020 (2016)] based on a mutual information criterion for EPR steering.
 \end{abstract}



\maketitle


Several no-go theorems \cite{nogo} and nonclassical correlations \cite{RevModPhys.81.865,RMP2012Vedral,RMP2014bell,RMP2020steer} have been presented
to describe the difference between the quantum and classical worlds.
These properties have the advantage of  quantum  information processes
and also forbid  quantum counterparts of  some classical tasks \cite{Book}.

The no-cloning theorem
presented in 1982 states that
no physical process can perfectly duplicate any pure state \cite{clone}.
It has been extended to the case of mixed states and resulted in the no-broadcasting theorem \cite{PRL1996Barnum}.
Several other no-go theorems have been proposed sequentially,
including
no-deletion theorem \cite{AKP},
no-flipping,
no-self replication \cite{AKP1},
and the most recent  no-masking theorem \cite{PRL2018mask}.
Alternatively,
several versions of imperfect quantum cloning machines that can clone an unknown state with high fidelity have been designed \cite{BH}.
These machines could be applied in eavesdropping attacks on the protocols of quantum cryptography \cite{RMP2005cloning}.
In addition,
the no-cloning theorems of quantum properties in quantum systems or their behaviors under cloning operations
have recently attracted the attention of researchers.
For instance,
no operation can perfectly clone entanglement
and simultaneously preserve separability \cite{PRA2004cloning},
and it is impossible to clone the coherence of an arbitrary quantum state \cite{PRA2021impossibility}.

The phenomenon of quantum steering,
termed by Schr\"{o}dinger \cite{SCat},
reveals that
Alice can project Bob's system into different states by performing different measurements on her system
when they share a correlated state.
Even if Alice is untrusted by Bob,
some states can be used to convincingly demonstrate this phenomenon,
which is an asymmetric quantum correlation lying between Bell nonlocality and entanglement \cite{PRL2007Steering}.
EPR steering is an important resource in  one-sided device-independent quantum information tasks   \cite{RMP2020steer,PRA2012OneSDIQKD,JPA2014quantum}.
The concept of steering  also acts as a critical research tool for theoretical issues in quantum information,
such as the Peres conjecture   \cite{Peres1999,NP2014,PhysRevLett.113.050404},
Gisin's theorem \cite{Gisin1991,SR2015Chen},
and the construction of local hidden-variable models for entangled states \cite{Arxiv2015LHV, PRL2016Algorithmic,LHVGHZ}.

Here, we raise a simple question of
 \emph{whether it is possible to clone quantum steering in a  quantum state}.
A closely related issue has been recently explored \cite{npj2016},
in which one-half of a maximally entangled state was cloned by an imperfect quantum cloning machine.
EPR steering verified  by a mutual information criterion \cite{PRA2015MI}
cannot be simultaneously observed in the two copies.

This study
distinguishes  the concepts of \emph{quantum steering}  and \emph{EPR steering}.
The former establishes a correspondence between each of  Alice's measurement outcomes and a conditional state of Bob's system \cite{PRL2014Ellips,PHDSteer2002,PRA2016steer}.
It is characterized by the set of Bob's unnormalized conditional states,  referred to as an  {assemblage} \cite{PRA2016steer}.
%
The operational definition presented by
Wiseman \emph{et al.} \cite{PRL2007Steering},
a bipartite state exhibits  EPR steering if and only if  its assemblage cannot be reproduced by an LHS model.

When cloning quantum steering,
a perfect duplicate should result in the same assemblage as in the original state.
We focus on operations on the steered part.
 A necessary and sufficient condition for perfect cloning
 is that the initial state is of zero discord from the steered part to the measuring part.
We also revisit the process studied in Ref. \cite{npj2016}
and focus on the case of qubits.
%
EPR steering can simultaneously appear in both of the copy subsystems,
which contradicts the conclusion of \emph{no-cloning of quantum steering (EPR steering)} in Ref. \cite{npj2016}.
The EPR steering in the two clones is bounded by an inequality
derived from  the optimal fidelity of the two copied assemblages.
This study measures EPR steering using the average length of steered Bloch vectors \cite{HGG2022},
which is $1/2$ on the steerable boundary for the two-qubit states appearing in the process \cite{jevtic2015einstein,nguyen2016necessary,ZhangPRA2019}.

\emph{ Steering No-cloning Principle.-- }
We begin by introducing some notions of steering and its cloning.
Let $\rho_{AB}$ denote the quantum state prepared  by Alice.
She keeps part A and sends B to Bob.
On her part,  Alice makes a positive operator-valued measurement,
described by operators $\{ \varPi^x_a \}$,
with  $ \varPi^x_a  \geq 0 $ and $ \sum_a \varPi^x_a  = \openone $,
where $x$ denotes the observable and $a$ its outcome.
  %
Alice's measurement projects B onto the (unnormalized) conditional state
\begin{eqnarray}\label{Assem}
\rho_{a|x}^{B}=\tr_A [(\varPi^x_a \otimes \openone )\rho_{AB}],
\end{eqnarray}
 where $\tr_A$ is the partial trace over subsystem  A.
The probability of Alice's outcome is given by $P_{a|x}=\tr \rho_{a|x}^{B}$
and
the postmeasured state can be normalized as
 $\bar{\rho}_{a|x}^{B}=\rho_{a|x}^{B}/P_{a|x}$.
%
The assemblage of $\rho_{AB}$ is defined as the collection of $\{ \rho_{a|x}^{B} \}_{a,x} $,
which has the properties of $\rho_{AB}$ under Alice's measurements.

When cloning the quantum steering of $\rho_{AB}$,
 B goes through a cloning machine before being received by Bob.
The cloning machine performs a physical operation, $\mathcal{E}_{BC}$, on B and another system C in a standard quantum state $|0\rangle_C$,
  creating a tripartite state
\begin{eqnarray}\label{rhoabc}
\tilde{\rho}_{ABC}=\mathcal{E}_{BC}(\rho_{AB} \otimes |0\rangle_C \langle 0|).
\end{eqnarray}
After the operation, part B is received by Bob, and C is  received by a third observer, Charlie.
Then, the two states, shared by Alice and Bob, and Alice and Charlie, are given as follows:
\begin{eqnarray}\label{copies}
\tilde{\rho}_{AB}=\tr_{C}  \tilde{\rho}_{ABC},\ \ \ \tilde{\rho}_{AC}=\tr_{B}  \tilde{\rho}_{ABC}.
\end{eqnarray}
 For perfect cloning of quantum steering,
 the two states, $\tilde{\rho}_{AB}$ and $\tilde{\rho}_{AC}$,
 must result in  the same correspondence between Alice's measurement outcomes and the conditional states as the original state,
\begin{eqnarray}\label{clonecond}
\tilde{\rho}_{a|x}^{B}=\rho_{a|x}^{B},\ \ \ \tilde{\rho}_{a|x}^{C}=\rho_{a|x}^{B},
\end{eqnarray}
where
$\tilde{\rho}_{a|x}^{B}=\tr_A [(\varPi^x_a \otimes \openone )\tilde{\rho}_{AB}]$
 and
$\tilde{\rho}_{a|x}^{C}=\tr_A [(\varPi^x_a \otimes \openone )\tilde{\rho}_{AC}]$.
This is equivalent to the cloning of the  assemblage of $\rho_{AB}$,  as Alice's local measurement and $\mathcal{E}_{BC}$ commute.
%
The motivation for cloning quantum steering is to expand the one-to-one quantum information tasks based on steering to the one-to-many case.
 Alice can act as the cloning machine's implementer
 after preparing the state $\rho_{AB}$,
expecting to establish the same relationship with Charlie as the one with Bob.
Because Alice has a \emph{prior} knowledge of the form of $\rho_{AB}$,
$\mathcal{E}_{BC}$ is allowed to be $\rho_{AB}$ dependent,
which is a different assumption from the task in the no-cloning theorem.

It is obvious that
 no physical process can perfectly clone the quantum steering of  an entangled pure state.
The simplest example is one of the Bell states of two qubits, which is given as follows:
\begin{eqnarray}\label{BellState}
|\varPhi^+\rangle_{AB}=\frac{1}{\sqrt{2}}\bigr(|0\rangle|0\rangle+|1\rangle|1\rangle\bigr)
=\frac{1}{\sqrt{2}}\bigr(|\!+\!\rangle|\!+\!\rangle+ |\!-\!\rangle|\!-\!\rangle\bigr),
\end{eqnarray}
with $|\pm\rangle=(|0\rangle\pm|1\rangle)/\sqrt{2}$.
For Alice's von Neumann measurement on the basis $\{ |0\rangle_A, |1\rangle_A \} $,
the cloning operation must satisfy
$\mathcal{E}_{BC}( |0\rangle_B |0\rangle_C ) = |0\rangle_B |0\rangle_C $
and
$\mathcal{E}_{BC}( |1\rangle_B |0\rangle_C ) = |1\rangle_B |1\rangle_C $,
whereas  the basis $\{ |\pm\rangle_A \} $ requires $\mathcal{E}_{BC}( |\pm\rangle_B |0\rangle_C ) = |\pm\rangle_B |\pm\rangle_C$.
The  existence of such $\mathcal{E}_{BC}$ is excluded by the no-cloning theorem.

The question now becomes  whether the quantum steering in a general (or say mixed) state can be perfectly cloned.
We show that a perfect cloning exists  if and only if the original states is of the following form:
\begin{eqnarray}\label{D0}
\rho_{AB} = \sum_{j} p_j \rho^A_j \otimes | \beta_j\rangle_B  \langle \beta_j |
\end{eqnarray}
where $\{| \beta_j\rangle_B \}$ is a set of orthonormal basis for B,
$\rho^A_j$ are the  states of part A,
and $p_j$ are nonnegative numbers such that $ \sum_{j} p_j =1$.
These states have a zero discord from B to A \cite{RMP2012Vedral} and
 are called quantum-incoherent states \cite{Chitambar2016,Matera2016coherent,Streltsov2017towards} in the resource theory of quantum coherence.

The perfect operation for the states (\ref{D0})  can be simply designed as any unitary operator, $U_{BC}$, which is consistent with
 $\mathcal{E}_{BC}( |\beta_j\rangle_B |0\rangle_C ) =U_{BC}  |\beta_j\rangle_B |0\rangle_C = |\beta_j\rangle_B |\beta_j\rangle_C $.
It is direct to verify that the conditions (\ref{clonecond}) are fulfilled for arbitrary measurement and outcome.
The necessity can be proved  using the no-broadcasting theorem \cite{PRL1996Barnum}.
Let the dimensions of the local Hilbert spaces of A be $d$,
 and $\lambda_k$ with $k=1,...,d^2-1$ be generators of $SU(d)$,
which satisfy  $\lambda_k^\dag =\lambda_k$, $\tr \lambda_k =0$ and $ \tr (\lambda_k \lambda_l) =2 \delta_{kl}$.
An original state can always be decomposed as follows:
\begin{eqnarray}
\rho_{AB} = \frac{1}{d}(\openone \otimes \rho_B +  \sum_{k} \lambda_k \otimes \eta_k )
\end{eqnarray}
 where $ \rho_B= \tr_A \rho_{AB}  $ is the reduced state of B
 and $\eta_k =  \tr_A   (\lambda_k  \otimes \openone  \rho_{AB}) d/2 $ is $d^2-1$ Hermitian operators.
We define $d^2-1$ pairs of binary measurement operators
as $ \varPi^k_a   =\frac{1}{2} (\openone + a  \Delta \lambda_k)$
with $a=\pm1$ and $\Delta \in (0, 1/\sqrt{2})$.
By substituting them into the protocol described by Eqs. (\ref{Assem})-(\ref{copies}),
one can find that the
conditions (\ref{clonecond})
require broadcasting the set of states $\{\rho_{a|k}^{B} =  \frac{1}{2}\rho_B + \frac{1}{d} a  \Delta  \eta_k \}$.
The no-broadcasting theorem \cite{PRL1996Barnum} indicates that
the task is viable if and only if $[\rho_{a|k}^{B} ,\rho_{a'|k'}^{B} ]=0$ for arbitrary $\{a,k;a',k'\}$.
It is equivalent to $[\rho_B ,\eta_k ]=[\eta_k ,\eta_{k'}] =0$.
By denoting  the common eigenstates of $\rho_B$ and all $\eta_k$ as $\{| \beta_j\rangle_B \}$,
one can write  $\rho_{AB}$ in the form (\ref{D0}).

\emph{Imperfect Cloning.--}
We now consider an imperfect realization of $\mathcal{E}_{BC}$  using Cerf's cloning machine \cite{cerf1998information,cerf2000pauli,npj2016}.
 We focus on the systems of qubits with the original state, $\rho_{AB}=|\varPhi^+\rangle_{AB} \langle \varPhi^+|$, and von Neumann measurements on Alice's part.
%
Chiu \emph{et al.} \cite{npj2016} have investigated the same process in the case of qudits
and proved EPR steering  using a sufficient criterion based on  mutual information \cite{PRA2015MI}.
However, we obtained an opposite conclusion to theirs since
the EPR steering in this study has a computable necessary and sufficient condition \cite{jevtic2015einstein,nguyen2016necessary,ZhangPRA2019}.

\begin{figure}
\centering
\includegraphics[height=3.6cm]{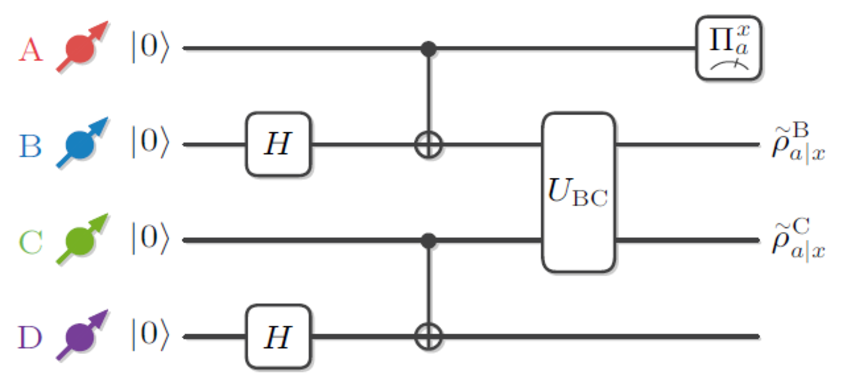}
\caption{Imperfect cloning of quantum steering.
Alice prepares the Bell state, $|\varPhi^+\rangle_{AB}$,
by applying a Hadamard gate and  a controlled-NOT gate on qubits A and B in standard states.
Charlie prepares $|\varPhi^+\rangle_{CD}$  using  the same operations on C and D.
He intercepts qubit B  sent by Alice to Bob and performs the unitary (\ref{UBC}) on  B and C.
He keeps C and routes B to Bob.
Then, Alice's measurement on A, labeled  $x$ and with outcome $a$,
projects B and C into  the (unnormalized) conditional states, $\tilde{\rho}_{a|x}^{B}$ and $\tilde{\rho}_{a|x}^{C}$.
 }
\label{CerfClone}
\end{figure}

In this section, we assume Charlie to be the cloning machine's implementer,
 acting as an eavesdropper that attacks the communication between Alice and Bob.
%
According to Cerf's protocol \cite{cerf1998information,cerf2000pauli}
(Fig. \ref{CerfClone}),
Charlie entangles C and an auxiliary qubit, D, into the Bell state, $|\varPhi^+\rangle_{CD}$,
and performs a unitary
\begin{eqnarray}\label{UBC}
 U_{BC} =v_0 \openone +\sum_{k=1}^3 v_k \sigma_{k}\otimes\sigma_{k},
\end{eqnarray}
on qubit B and C, with $\sigma_{k}$ being the Pauli matrices and $\sum_{k=0}^3 |v_k|^{2}=1$.
The total state becomes
\begin{eqnarray}
|\tilde{\varOmega} \rangle  &=& (\openone \otimes U_{BC} \otimes \openone )  |\varPhi^{+}\rangle_{AB}|\varPhi^{+}\rangle_{CD} \nonumber \\
&=& (  v_0 |\varPhi^{+}\rangle|\varPhi^{+}\rangle+ v_3|\varPhi^{-} \rangle|\varPhi^{-}\rangle  \nonumber \\
                        &\ & \  +v_1|\varPsi^{+}\rangle |\varPsi^{+}\rangle+v_2|\varPsi^{-}\rangle |\varPsi^{-}\rangle)_{ABCD},
\end{eqnarray}
 with the Bell states
$|\varPhi^{\pm}\rangle=\frac{1}{\sqrt{2}}(|00\rangle\pm|11\rangle) $ and  $|\varPsi^{\pm}\rangle=\frac{1}{\sqrt{2}}(|01\rangle\pm|10\rangle)$.
Then, the tripartite state is $\tilde{\rho}_{ABC} = \tr_{D} (|\tilde{\varOmega}\rangle \langle\tilde{\varOmega} |)$.
The form of $|\tilde{\varOmega}\rangle$ is preserved by an exchange between B and C,
  with the parameters  being replaced by
\begin{eqnarray}
&{v_0'}=\frac{1}{2}(v_0+v_3+v_1+v_2), \
{v_3'}=\frac{1}{2}(v_0+v_3-v_1-v_2), \   \nonumber \\
&{v_1'}=\frac{1}{2}(v_0-v_3+v_1-v_2), \
{v_2'}= \frac{1}{2}(v_0-v_3-v_1+v_2). \  \nonumber
\end{eqnarray}
Therefore,
 the states
shared by Alice and Bob, $\tilde{\rho}_{AB}$,
and Alice and Charlie, $\tilde{\rho}_{AC}$, are two Bell diagonal states.

We define the fidelities of the cloning as follows:
\begin{eqnarray}
F_B= \sum_{x,a} q_x P_{a|x} F(\bar{\rho}_{a|x}^{B},\bar{\tilde{\rho}}_{a|x}^{B} ),\\
F_C= \sum_{x,a} q_x P_{a|x}F(\bar{\rho}_{a|x}^{B},\bar{\tilde{\rho}}_{a|x}^{C} ),
\end{eqnarray}
where $F(\rho_1,\rho_2) =\tr (\sqrt{\rho_1} \rho_2 \sqrt{\rho_1})$
and $q_x$ is the probability of observable $x$ in Alice's local measurements.
Here, $\bar{\tilde{\rho}}_{a|x}^{B}=\tilde{\rho}_{a|x}^{B}/P_{a|x} $ and $\bar{\tilde{\rho}}_{a|x}^{C}=\tilde{\rho}_{a|x}^{C}/P_{a|x} $ are normalized conditional states of two copies,
with their probabilities $\tr \tilde{\rho}_{a|x}^{B} = \tr \tilde{\rho}_{a|x}^{C} = P_{a|x}$.
The values of $F_B$ and $F_C$ are the average fidelities of two copies of a normalized conditional state compared with the ones of  the original state.
%
Alice's measurement operator can be written as $\varPi^x_a  = \frac{1}{2}(\openone+ a  \vec{x} \cdot \vec{\sigma}) $,
with $\vec{x}$ being a unit vector and denoting the measurement direction.
Let us assume $\vec{x}$ is completely randomized on the unit sphere.
We choose $v_0$ to be a nonnegative real coefficient, without a loss of generality.
The fidelities can be obtained by straightforward calculation as follows:
 \begin{eqnarray}
F_{B}= \frac{1}{3}(1+2 v_0^{2}) , \ \ \
F_{C}= \frac{1}{3}(1+2 {v_0'}^{2}).
\end{eqnarray}
Here, $F_C$ is maximized when $v_k$ are nonnegative real numbers for a fixed $\tilde{\rho}_{AB}$ or equivalent $\{|v_{k}|\}$.
Further, the maximum of $F_{C}$ occurs at $v_{k\neq0} =\sqrt{\frac{1}{3}(1-v_0^2)} $ for a fixed $F_{B}$ or equivalent $v_0$.
Then, $F_{B}$ and $F_{C}$  saturate the inequality \cite{cerf2000pauli},
%
 \begin{eqnarray}\label{noclone}
(1-F_{B})+(1-F_{C})+\sqrt{(1-F_{B})(1-F_{C})} \geq \frac{1}{2},
\end{eqnarray}
which bounds the two fidelities
when the directions of $\vec{x}$ are uniformly distributed on the unit sphere.
This type of no-cloning inequality arises as a condition of optimality in various studies on universal cloning machines.

\begin{figure}
 \centering
 \includegraphics[height=6cm]{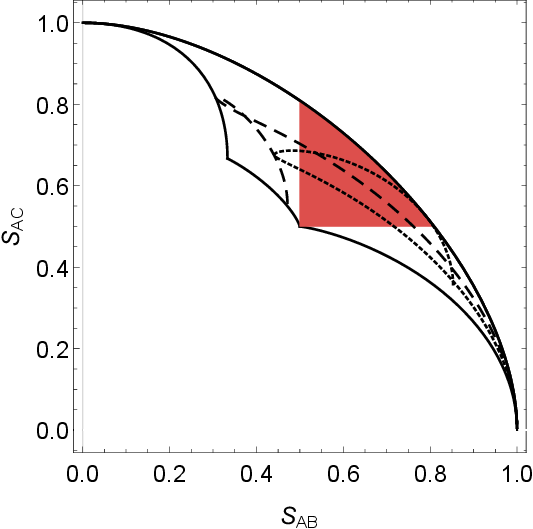}
 \caption{
The top solid curve
 (with the  nonnegative real coefficients $v_1=v_2=v_3\leq v_0/\sqrt{3}$)
 and  the two axes surround the areas of $\tilde{\rho}_{AB}$ and $\tilde{\rho}_{AC}$.
The solid curves show the boundary of the results with nonnegative real $v_k$
 and the three sections of the lower bound are obtained when $(1)\ \ v_1=v_2=v_3> v_0/\sqrt{3}$, $(2)\ \ v_1=v_2,v_3=0$,  and $(3)\ \ v_2=v_3=0$, from left to right.
The dashed and dotted curves show the results with $ v_1=v_2=4v_3 $ and $ v_1=v_2=v_3/4$, respectively.
The double steerable region with $S_{AB}>\frac{1}{2} \cap S_{AC}>\frac{1}{2}$ is colored.
}
\label{figS}
\end{figure}

EPR steering, with the operational definition provided by Wiseman \textit{et al.} \cite{PRL2007Steering}, in a Bell diagonal state, $\rho$, can be demonstrated by its correlation matrix $T$ as follows:
\begin{equation}\label{ST}
S=   \int \frac{1}{4 \pi}  |T \vec{x}| d \vec{x},
\end{equation}
where the integral is over the unit sphere and $d \vec{x}$ is the surface element.
Here, $T$ is a $3\times3$ diagonal matrix with the diagonal elements  $T_k=\tr (\sigma_{k}\otimes\sigma_{k} \rho)$,
and $S$ measures the  average length of  its steered Bloch vectors for a completely random measurement direction.
The Bell diagonal  state $\rho$ is steerable if and only if $S>1/2$.
We adopt the values of $S$ for $\tilde{\rho}_{AB}$ and $\tilde{\rho}_{AC}$, $S_{AB}$, and $S_{AC}$, as the figure of merit of their EPR steering.
Their correlation matrices have the elements $|T_k|=|2 (v_0^2 - |v_k|^2)-1| $ and $|T_k'|=|2 ({v_0'}^2 - {|v_k'|}^2)-1|$.

 Figure \ref{figS}
shows the regions of $S_{AB}$ and $S_{AC}$
 for the general case,
 the case with nonnegative real $v_k$,
 and the two steerable copies, $\tilde{\rho}_{AB}$ and $\tilde{\rho}_{AC}$.
When the coefficients  $v_k$ are nonnegative real numbers,  for fixed $v_1 / v_3$ and $v_2 / v_3$,
the curves for $S_{AB}$ and $S_{AC}$  always intersect the double steerable region as $v_0$ decreases from $1$ to $0$,
 except when two values of $v_k$ are zero.
In addition,
the upper bound of the general region is reached by the coefficients $v_1=v_2=v_3\leq v_0/3$,
which also saturate the no-cloning inequality (\ref{noclone}).
Consequently, this results in
 an inequality for the EPR steering in  the current protocol as follows:
 \begin{eqnarray}
(1\!- \!S_{AB})+(1\!- \!S_{AC})+\sqrt{(1\!-\!S_{AB})(1\!-\!S_{AC})} \geq 1.\
\end{eqnarray}
This can be regarded as a  conservation law of EPR steering when equality holds.


Our results contradict the \emph{no-cloning} conclusion in Ref. \cite{npj2016}, as EPR steering can be simultaneously observed in the two copies.
Consequently, the two quantum information applications proposed in Ref. \cite{npj2016} require further investigation.
 First,
implement the quantum key distribution,
in which channels are secured against cloning-based attacks by ruling out false steering,
the EPR steering between a sender (Alice) and a receiver (Bob) must exceed some thresholds.
In the process described in this study (Fig. \ref{figS}),
when $S_{AB} > (1+\sqrt{5})/4 \simeq 0.809$, and
Alice and Bob can be convinced that
no EPR steering was formed between an eavesdropper (Charlie) and Alice  using  a cloning machine for coherent attacks.
Second,
a similar requirement should be satisfied
to guarantee an accurate implementation of a quantum computing system by observing EPR steering under uncharacteristic measurements and cloning-based attacks.
How Alice and Bob simply verify whether their EPR steering  has been partially cloned by any eavesdropper remains unclear.

In summary,
we have investigated the task of cloning quantum steering in a quantum state.
Perfect cloning is feasible if and only if the initial state is of zero discord.
We also demonstrate that
EPR steering in a Bell state can be partially cloned onto two copies using Cerf's quantum cloning machine,
which contradicts the conclusion of \emph{no-cloning of quantum steering (EPR steering)} in Ref. \cite{npj2016}, which was confirmed by a mutual information criterion.

Extending our results in numerous directions would be intriguing.
First, one can try deriving an operational interpretation of discord in the framework of cloning states or quantumness.
A logical question is whether our results on imperfect cloning of the Bell state  can be  applied to higher-dimensional systems.
%
The key technology at this point is to derive more effective  criteria for EPR steering in multidimensional systems
since our results are based on the necessary and sufficient conditions for Bell diagonal states.
It would also be interesting to consider a unified view of the imperfect cloning of steering, teleportation, and multipartite maskers
since these protocols have a similar formalisms.
 The relationship between the latter two has recently been established \cite{shang2021quantum}.
We particularly hope that the recent advances in teleportation \cite{luo2019quantum,hu2020experimental} and quantum masking \cite{liu2021photonic,zhang2021experimental} will enable the
imperfect cloning of steering to be implemented in laboratories.

%

\emph{Acknowledgment.--} This work was supported by the NSF of China (Grants No. 11675119, No. 11575125, and No. 11105097).
We are grateful for comments from the anonymous referees.

\bibliography{CloneSteer}

\end{document}